\documentclass[pra,twocolumn,amsmath,amssymb,showpacs,floatfix]{revtex4}

\usepackage{graphicx}
\usepackage{color}

\pdfoutput=1

\newcommand{\Ham}{\hat{\mathcal{H}}}

\newcommand{\refe}[1]{Eq.~(\ref{#1})}
\newcommand{\reff}[1]{Fig.~\ref{#1}}
\renewcommand{\vec}[1]{\mathbf{#1}}
\newlength\figurewidth
\begin{document}

\title{Exotic roton excitations in quadrupolar Bose-Einstein condensates}
\begin{abstract}
We investigate the occurrence of rotons in a quadrupolar Bose-Einstein condensate confined to two dimensions. Depending on the particle density, the ratio of the contact and quadrupole-quadrupole interactions, and the alignment of the quadrupole moments with respect to the confinement plane, the dispersion relation features two or four point-like roton minima, or one ring-shaped minimum.
We map out the entire parameter space of the roton behavior and identify the instability regions.
We propose to observe the exotic rotons by monitoring the  characteristic density wave dynamics  resulting from a short local perturbation, and discuss the possibilities to detect the predicted effects in state-of-the-art experiments with ultracold homonuclear molecules.
\end{abstract}  
\author{M.~Lahrz$^{1,2}$, M.~Lemeshko$^{3}$, L.~Mathey$^{1,2,4}$}
\affiliation{
\mbox{$^{1}$Zentrum f\"ur Optische Quantentechnologien, 
Universit\"at Hamburg, 22761 Hamburg, Germany}\\
\mbox{$^{2}$Institut f\"ur Laserphysik, Universit\"at Hamburg, 22761 Hamburg, Germany}\\
\mbox{$^{3}$IST Austria (Institute of Science and Technology Austria), Am Campus 1, 3400 Klosterneuburg, Austria}\\
\mbox{$^{4}$The Hamburg Centre for Ultrafast Imaging, Luruper Chaussee 149, Hamburg 22761, Germany}
}
\date{\today}

\maketitle


\section{Introduction}
Roton excitations, inherent to non-ideal superfluids with finite-range interactions, were first discussed in the seminal works of Landau~\cite{Landau1941}, Feynman~\cite{Feynman1957,Feynman1972}, and Bogoliubov~\cite{Bogoliubov1947} on the theory of liquid ${}^4\mathrm{He}$,  for which indeed a local minimum in the dispersion was observed. 
 The existence of such a minimum in a non-monotonic dispersion is in itself an intriguing scenario. Additionally, it can be seen as the precursor of a non-trivial order, as the roton softens. As the magnitude of the dispersion at the minimum approaches zero, the system tends to develop an instability, typically towards density wave order. Furthermore, it was later speculated that this instability not necessarily results in a suppression of superfluidity, meaning that the softening of the roton mode would give rise to a supersolid~\cite{Chester1970,Leggett1970,Schneider1971}. While the occurrence of a supersolid phase in helium has been a subject of an active debate for over fifty years~\cite{Kim2004,Kuklov2011}, no conclusive experimental evidence of supersolidity has been found yet~\cite{Kim2012,Mi2014}.

As opposed to liquid helium, whose properties can  be controlled primarily through global, thermodynamic quantities, such as pressure and temperature, ultracold quantum gases allow for a versatile tunability of  the microscopic Hamiltonian. As an example, a roton instability was predicted to arise in Bose-Einstein condensates (BECs) of dipolar particles confined to one- and two-dimensional geometries~\cite{Santos2003,Wilson2008,Lahaye2009,Bisset2013}, as well as in a BEC of nonpolar atoms in the presence of an intense laser light~\cite{ODell2003} or Rydberg dressing~\cite{Henkel2010}.

Recently we introduced ultracold quantum gases of quadrupolar particles as a perspective platform for studying many-body phenomena~\cite{Bhongale2013, Huang2014, Lahrz2014}. Quadrupolar particles, such as ultracold homonuclear dimers, are prone to chemical reactions occurring for dipolar molecules~\cite{deMiranda2011}. On the other hand, since anisotropic quadrupole-quadrupole interactions occur in the molecular ground state, the coherence time is not disturbed by spontaneous emission due to scattering of laser photons~\cite{ODell2003,Henkel2010}. Finally, although the quadrupole-quadrupole interactions are of shorter range compared to the dipole-dipole ones~\cite{Lahaye2009}, particles possessing electric quadrupole moments, such as $\mathrm{Cs}_{2}$~\cite{Herbig2003} or $\mathrm{Sr}_{2}$~\cite{Stellmer2012,Reinaudi2012}, are readily available in experiments at higher densities compared to dipolar species. Among the exciting properties of the quadrupole-quadrupole interactions is their peculiar anisotropy, which, combined with their broad tunability,  paves the way to observing novel quantum phases in ultracold experiments~\cite{Bhongale2013, Huang2014,Lahrz2014}.

In this contribution we investigate the occurrence of roton instabilities due to the interplay of quadrupole-quadrupole and contact interactions and explore the possibilities to detect the fingerprints of rotons in modern experiments with ultracold molecules. The paper is organized as follows: In Sec.~\ref{sec:theory} we introduce the system geometry and two-body interactions and sketch the derivation of the excitation spectrum in the framework of Bogoliubov's theory. A discussion of the stabilization criteria, Sec.~\ref{sec:stabilization}, is followed by the classification and occurrence of the rotons in the parameter space, Sec.~\ref{sec:rotons}. In Sec.~\ref{sec:realspace} we describe the dynamics of a quadrupolar BEC following a  short, local perturbation of the density, which can serve as an experimental detection tool for the roton instability in a BEC of homonuclear molecules. Finally, we conclude in Sec.~\ref{sec:conclusion}.


\section{Quadrupolar condensates}\label{sec:theory}
We investigate a two-dimensional, zero temperature BEC of density $n$. The bosons are interacting via quadrupole-quadrupole interactions (QQI) as well as contact interactions which can be tuned independently.
For this system, we derive the Bogoliubov spectrum and identify the roton excitations which can be stabilized in this system. The most interesting scenario of four roton minima is achieved by a competition of QQI and contact interaction and for a large alignment angle $\theta_{F}$.

%
\subsection{Quadrupole-quadrupole interaction}
\begin{figure}
\begin{center}
\includegraphics[width=\figurewidth]{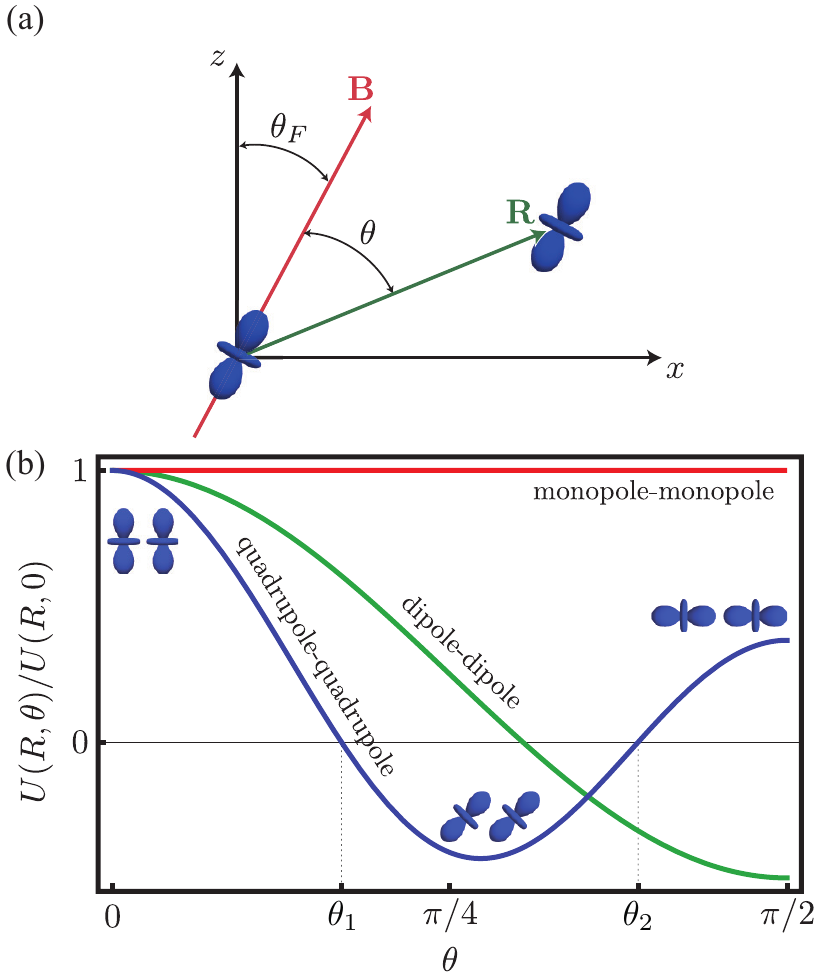}
\caption{(Color online)
(a) Two quadrupoles, aligned along a magnetic field $\vec{B}$, are separated by a vector $\vec{R}$. The resulting quadrupole-quadrupole interaction, \refe{QQI},  is a function of their distance, $\left|\vec{R}\right|$, and the angle, $\theta$, between $\vec{R}$ and $\vec{B}$. If the tilting is along the $x$-axis, $\theta$ can be expressed via $\theta_{F}$ and $\vec{R}$.
(b) Quadrupole-quadrupole interaction as a function of $\theta$ in the range of $0$ to $\pi/2$, where it changes from repulsive to attractive and back. The angles at which the interaction switches its sign are labeled as $\theta_{1}$ and $\theta_{2}$. For comparison, we also show the angular dependence of the dipole-dipole interaction, which has only one zero-crossing, and the monopole-monopole interaction, which is constant. Adapted from Ref.~\cite{Lahrz2014}.}
\label{fig:geometry}
\end{center}
\end{figure}

We consider a  system of quadrupoles similar to the one described in Ref.~\cite{Lahrz2014}. However, here the particles are bosonic rather than fermionic. The interaction between two particles having a quadrupole moment $q$ separated by the distance vector $\vec{R}$ and aligned via an external magnetic field $\vec{B}$ is
\begin{align}\label{QQI}
U{\left(\vec{R}\right)}
&= C_{q}\frac{3-30\cos^2{\theta} + 35\cos^4{\theta}}{R^5}
\end{align}
where $C_{q} = 3q^{2}/\left(64\pi\varepsilon_{0}\right)$, $R = \left|\vec{R}\right|$, and $\theta$ the angle between $\vec{B}$ and $\vec{R}$, cf. \reff{fig:geometry} (a). The QQI changes its sign twice, at the angles $\theta_{1} \equiv \arccos{\sqrt{(15+2\sqrt{30})/35}} \approx 0.533$ and  $\theta_{2} \equiv \arccos{\sqrt{(15-2\sqrt{30})/35}} \approx 1.224$. The interaction is repulsive for $\theta \in \left[0,\theta_{1}\right)$ and $\theta \in \left(\theta_{2},\pi/2\right]$, and attractive for $\theta \in \left(\theta_{1},\theta_{2}\right)$.  In \reff{fig:geometry} (b) we show the angular dependence of the quadrupole-quadrupole interaction (QQI), in comparison to a dipole-dipole interaction, $U \propto 1-3\cos^{2}{\theta}$, and a monopole-monopole interaction, $U \propto 1$.

%

We consider a quasi-2D geometry, in which the motion of the particles in the $z$-direction is confined by a harmonic potential, $V_{\mathrm{c}} = m\omega_{z}^{2}z^{2}/2$, where $m$ is the particle mass and $\omega_{z}$ the oscillator frequency. The oscillator length is $\lambda_{z} = \sqrt{\hbar/\left(m\omega_z\right)}$. The 2D limit is achieved for $\hbar\omega_{z} \gg \mu$, where $\mu$ is the chemical potential of the system. In this limit, only the spatial ground state is occupied and we factorize the single-particle operator as $\hat{\Psi}{\left(\vec{R}\right)} = \hat{\psi}{\left(\vec{r}\right)}\chi{\left(z\right)}$ where $\chi{\left(z\right)}$ reads as 
\begin{align}\label{eq:chiz}
\chi{\left(z\right)} &= \frac{1}{\left(\pi\lambda_{z}^{2}\right)^{1/4}} \exp{\left(-\frac{z^2}{2\lambda_{z}^2}\right)}\,.
\end{align}
As discussed in \cite{Lahrz2014}, we integrate out the $z$-component, which leads to an effective 2D potential $U_{\mathrm{2D}}{\left(\vec{r}\right)}$ given by
\begin{align}\label{eq:U2D}
U_{\mathrm{2D}}(\vec{r})
&= \frac{1}{\left(2\pi\lambda_{z}^{2}\right)^{1/2}}\int{\mathrm{d}{z}U{\left(\vec{R}\right)}\exp{\left(-\frac{z^2}{2\lambda_{z}^2}\right)}}\,.
\end{align}
The full analytical expression for $U_{\mathrm{2D}}{\left(\vec{r}\right)}$ is given in Appendix \ref{sec:Fouriertrafo}. For $\left|\vec{r}\right|\gg \lambda_{z}$, $U_{\mathrm{2D}}{\left(\vec{r}\right)}$ approaches the $\sim 1/r^{5}$ behavior of the bare interaction. For $\left|\vec{r}\right|\ll \lambda_{z}$, the $\sim 1/r^{5}$ divergence is suppressed to a lower power, which makes $U_{\mathrm{2D}}{\left(\vec{r}\right)}$ sufficiently well-behaved on short scales, so that no additional short range cut-off has to be introduced, see Ref.~\cite{Lahrz2014}. 

The Fourier transform of this interaction is given by
\begin{align}\label{eq:V2D}
V_{\mathrm{2D}}{\left(k,\beta\right)} &= \int{U_{\mathrm{2D}}(\vec{r})\mathrm{e}^{-\mathrm{i}\vec{r}\cdot\vec{k}}\mathrm{d}{\vec{r}}}\,,
\end{align}
where we expressed the momentum $\vec{k}$ in terms of its absolute value, $k \equiv \left|\vec{k}\right|$, and the angle between the vector and the $x$-axis, $\beta \equiv \arg{\vec{k}}$. The full analytic solution is sketched in Appendix \ref{sec:Fouriertrafo}.

\reff{fig:V2D} shows $V_{\mathrm{2D}}{\left(k,\beta\right)}$ for different tilting angles $\theta_{F}$. The competition between attractive and repulsive contributions in different regimes of momentum space leads to interesting quantum phases of quadrupolar systems, as discussed in Refs.~\cite{Bhongale2013,Huang2014}.
\begin{figure}
\begin{center}
\includegraphics[width=\figurewidth]{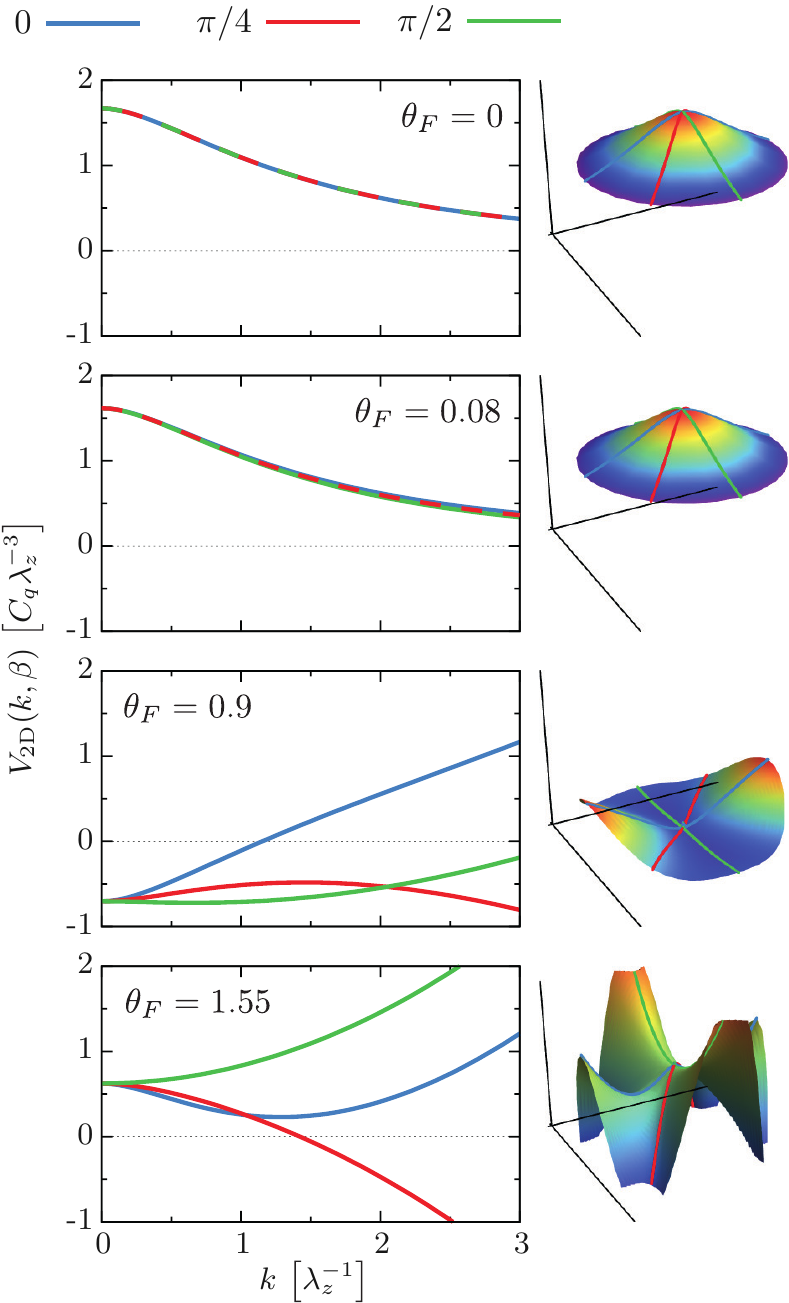}
\caption{Fourier transform of the quadrupole-quadrupole interaction, $V_{\mathrm{2D}}{\left(k,\beta\right)}$, for different values of the tilting angle $\theta_{F}$. Here, the quadrupoles are tilted along the $x$-axis. (a) For $\theta_{F} = 0$ the interaction is rotationally symmetric and repulsive for all momenta. For $\theta_{F} > 0$ the rotational symmetry is broken. (b) For $\theta_{F} = 0.08$, this symmetry breaking is barely visible on this scale. However, even for this angle the parameters can be tuned such that point-like rotons appear in the spectrum. (c) For $\theta_{F} = 0.9$ the interaction is attractive for small momenta. The anisotropy is clearly visible. (d) For $\theta_{F} = 1.55$ the low-momentum limit is repulsive again. As it is clearly visible on the right-hand side, there are four distinct directions in momentum space for which the interaction is attractive. We make use of this feature to create a novel condensate state with four distinct roton minima.}
\label{fig:V2D}
\end{center}
\end{figure}

\subsection{Contact interaction}

In addition to the QQI, we account for a contact interaction between the particles, as described by $V_{\mathrm{contact}} = \frac{1}{2}g_{\mathrm{3D}}\int_{\mathbb{R}^{3}}{\mathrm{d}{\vec{R}}\hat{\Psi}^{\dagger}{\left(\vec{R}\right)}\hat{\Psi}^{\dagger}{\left(\vec{R}\right)}\hat{\Psi}{\left(\vec{R}\right)}\hat{\Psi}{\left(\vec{R}\right)}}$. Here, the interaction strength is given by $g_{\mathrm{3D}} = 4\pi\hbar^{2} a/m$, where $a$ is the s-wave scattering length. 
We project this interaction onto two dimensions, by analogy to the QQI of \refe{eq:U2D}, which results in the term $V_{\mathrm{contact}} = \frac{1}{2}g_{\mathrm{2D}}\int_{\mathbb{R}^{2}}{\mathrm{d}{\vec{R}}\hat{\psi}^{\dagger}{\left(\vec{r}\right)}\hat{\psi}^{\dagger}{\left(\vec{r}\right)}\hat{\psi}{\left(\vec{r}\right)}\hat{\psi}{\left(\vec{r}\right)}}$.  The effective 2D interaction strength $g_{\mathrm{2D}}$ is given by 
\begin{align}\label{eq:g2D}
g_{\mathrm{2D}} &= g_{\mathrm{3D}}\int_{\mathbb{R}}{\mathrm{d}{z}\left|\chi{\left(z\right)}\right|^{4}} = \frac{\sqrt{8\pi}\hbar^{2} a}{m\lambda_{z}}\,.
\end{align}
In experiment, this interaction strength can be controlled by either a Feshbach resonance~\cite{Courteille1998,Inouye1998}, or by changing the confinement length scale $\lambda_{z}$.

\subsection{Bogoliubov spectrum}\label{Bogspec}
We derive the spectrum of the system within the Bogoliubov approximation. 
The Hamiltonian of the system is 
\begin{align}
\Ham_{0}
&= \sum_{\vec{k}}{\frac{\hbar^{2}k^{2}}{2m}\hat{a}_{\vec{k}}^{\dagger}\hat{a}_{\vec{k}}}
+\frac{1}{2A}
\sum_{\vec{k},\vec{q},\vec{p}}{
\hat{a}_{\vec{k}+\vec{p}}^{\dagger}
\hat{a}_{\vec{q}-\vec{p}}^{\dagger}
V{\left(\vec{p}\right)}
\hat{a}_{\vec{q}}
\hat{a}_{\vec{k}}
}
\end{align}
where $\hat{a}_{\mathbf{k}}$ is the annihilation operator of mode $\vec{k}$ and the Fourier transform of the single particle operator $\hat{\psi}$, $\hbar$ is Planck's constant, and $A$ is the system area. The interaction $V{\left(\vec{k}\right)} = V_{\mathrm{2D}}{\left(k,\beta\right)} + g_{\mathrm{2D}}$ contains both the QQI, \refe{eq:V2D}, and the contact interaction, \refe{eq:g2D}. We perform a Bogoliubov transformation of the form 
 $\hat{b}_{\vec{k}} = u_{\vec{k}}\hat{a}_{\vec{k}} - v_{\vec{k}}\hat{a}_{-\vec{k}}^{\dagger}$ where the Bogoliubov functions are given by $u_{\vec{k}}^{2} = (\hbar\omega_{\vec{k}}+\frac{\hbar^{2}k^{2}}{2m}+nV{\left(\vec{k}\right)})/\left(2\hbar\omega_{\vec{k}}\right)$ and $v_{\vec{k}}^{2} = (-\hbar\omega_{\vec{k}}+\frac{\hbar^{2}k^{2}}{2m}+nV{\left(\vec{k}\right)})/\left(2\hbar\omega_{\vec{k}}\right)$, respectively. This results in a linearized Hamiltonian 
\begin{align}\label{eq:Ham0}
\Ham_{0} &= \hbar\omega_{0} + \sum_{\vec{k}\neq0}\hbar\omega_{\vec{k}}\hat{b}_{\vec{k}}^{\dagger}\hat{b}_{\vec{k}}
\end{align}
where the dispersion relation of the quasi-particles is 
\begin{align}\label{eq:epsk}
\omega_{\vec{k}}^{2} &= \left(\frac{\hbar k^{2}}{2m}\right)^{2} + \frac{nk^{2}}{m}\left(V_{\mathrm{2D}}{\left(k,\beta\right)} + g_{\mathrm{2D}}\right)\,.
\end{align}
Due to the anisotropy of the QQI, the dispersion relation depends not only on the absolute momentum, $k = \left|\vec{k}\right|$, but also on its direction, $\beta = \arg{\vec{k}}$.


\section{Stability of the condensate}\label{sec:stabilization}
  Our main goal is to identify the parameter regime in which the dispersion relation (\ref{eq:epsk}) is non-monotonic and displays one or several roton minima. Furthermore, the dispersion at the roton minimum can become imaginary indicating a roton instability, which is often a precursor of a new, non-trivial order of the system. For example, as shown in Ref.~\cite{Macia2012}, a roton instability of 2D dipolar Bose gases precedes the formation of a striped phase. Below, we identify the roton instabilities for a quadrupolar condensate. However, two other types of instabilities are present in the system. The first one occurs at large momenta and is due to the strongly attractive behavior of the QQI at the short range, and the second one belongs to small momenta and is accompanied by the collapse of the condensate. 

\subsection{Stability criterium at large momenta}

For large momenta, $k\to\infty$, the Fourier transform of the QQI reads
\begin{align}
V_{\mathrm{2D}}{\left(k\to\infty,\beta\right)} &= \frac{\sqrt{2\pi}}{12}\frac{C_{q}}{\lambda_{z}^{3}}\cos{\left(4\beta\right)}\sin^{4}{\left(\theta_{F}\right)}\lambda_{z}^{2}k^{2}\,.\label{QQIhighk}
\end{align}
Note that it scales as $\sim k^{2}$ and therefore does not become negligible compared to the kinetic energy, as opposed to the contact interaction $g_{\mathrm{2D}}$. 
  In order to further explore the consequences of this short-range instability, an improved description of the interactions on atomic scales would have to be given.
 
The term in \refe{QQIhighk} is non-zero for any $\theta_{F} \neq 0$.
 For $\cos{\left(4\beta\right)} = -1$ this term achieves its largest, negative value and competes with the kinetic part of \refe{eq:epsk}. As a result, the system becomes unstable for $k\to\infty$.  This can be expressed as an upper limit for the density,
\begin{align}\label{eq:stabkinf}
n{\left(\theta_{F}\right)} &\leq n_{\mathrm{c}}\sin^{-4}{\left(\theta_{F}\right)}\,,
\end{align}
where the critical density is defined as
\begin{align}
n_{\mathrm{c}} &\equiv \frac{3\hbar^{2}\lambda_{z}}{\sqrt{2\pi}C_{q}m}\,.
\end{align}
We note that, since $n_{\mathrm{c}} \propto \lambda_{z}$, strong confinement decreases the critical density.

As we demonstrate below, $n_{c}$ also defines the scale for the parameter regime in which rotons exist. 
In order to give a quantitative example, 
 we consider $\mathrm{Cs}_{2}$, with $q =  27.9\,\mathrm{a.u.}$~\cite{Byrd2011} and $m = 266\,\mathrm{a.u.}$ We assume a trapping frequency of $\omega_{z} = 10\,\mathrm{MHz}$ corresponding to $\lambda_{z} = 4.6\,{\mathrm{nm}}$. With these values, the critical density is $n_{c} = 522\,\mu$m$^{-2}$.
  For a molecule of the same mass with a larger quadrupole moment of, e.g., $q = 50\,\mathrm{a.u.}$ or $q = 100\,\mathrm{a.u.}$, we find $n_{c} = 162\,\mu$m$^{-2}$ and $n_{c} = 41\,\mu$m$^{-2}$, respectively. This indicates that the scenario considered in this contribution is relevant for current experiments. 


\subsection{Stability criterium for small momenta}
In addition to the short-range instability, the system can also undergo a collapse, which is characterized by an instability at small momenta.
In this limit, the analytic expression of the Fourier-transformed QQI reads
\begin{align}\label{eq:V2Db}
V_{\mathrm{2D}}{\left(k\to0\right)} &= \frac{\sqrt{2\pi}}{12}\frac{C_{q}}{\lambda_{z}^{3}}\left(3-30\cos^{2}{\left(\theta_{F}\right)}+35\cos^{4}{\left(\theta_{F}\right)}\right)\,.
\end{align}
We note that (\ref{eq:V2Db}) is independent of $\beta$ and $k$. The dispersion relation for small $k$ is given by $\omega_{k\to 0} = c_{s} k$ with the sound velocity
\begin{equation}
c_{s} \equiv \sqrt{\frac{n}{m}\left(V_{\mathrm{2D}}{\left(k\to0\right)} + g_{\mathrm{2D}}\right)}.
\end{equation}
Therefore, the system is stable if $V_{\mathrm{2D}}{\left(k\to0\right)} + g_{\mathrm{2D}} \geq 0$. The contact interaction can prevent collapse if it fulfills the requirement 
\begin{align}\label{eq:stabk0}
g_{\mathrm{2D}} \geq -\frac{\sqrt{2\pi}}{12}\frac{C_{q}}{\lambda_{z}^{3}}\left(3-30\cos^{2}{\left(\theta_{F}\right)}+35\cos^{4}{\left(\theta_{F}\right)}\right)\,.
\end{align}
Depending on $\theta_{F}$, the lower bound might be positive, which is the case for  $\theta_{F} \in\left(\theta_{1},\theta_{2}\right)$,  or negative. 

We introduce the relative interaction strength,
\begin{align}
\eta &\equiv -\frac{g_{\mathrm{2D}}}{V_{\mathrm{2D}}{\left(k\to0\right)}}\,,\label{etadef}
\end{align}
which depends on $\theta_{F}$ through $V_{\mathrm{2D}}{\left(k\to0\right)}$. The $\eta$ parameter takes the values from 0 to 1. $\eta = 0$ refers to zero contact potential. $\eta = 1$ corresponds to a vanishing speed of sound, $c_{s} = 0$, which indicates the onset of collapse.
We note, that in this representation for any $\eta$ the contact interaction is set to have an opposite sign with respect to $V_{\mathrm{2D}}{\left(k\to0\right)}$. If $V_{\mathrm{2D}}{\left(k\to0\right)}$ is repulsive (i.e. $\theta_{F} < \theta_{1}$ or $\theta_{F} > \theta_{2}$), $c_{s}^{2} \geq 0$ is ensured by setting $\eta \leq 1$, that is a smaller and attractive contact interaction. However, if $V_{\mathrm{2D}}{\left(k\to0\right)}$ is attractive ($\theta_{1} < \theta_{F} < \theta_{2}$), a larger repulsive contact interaction and thus $\eta \geq 1$ is required in order to avoid the collapse of the condensate.



\section{Rotons}\label{sec:rotons}
\begin{figure}
\begin{center}
\includegraphics[width=\figurewidth]{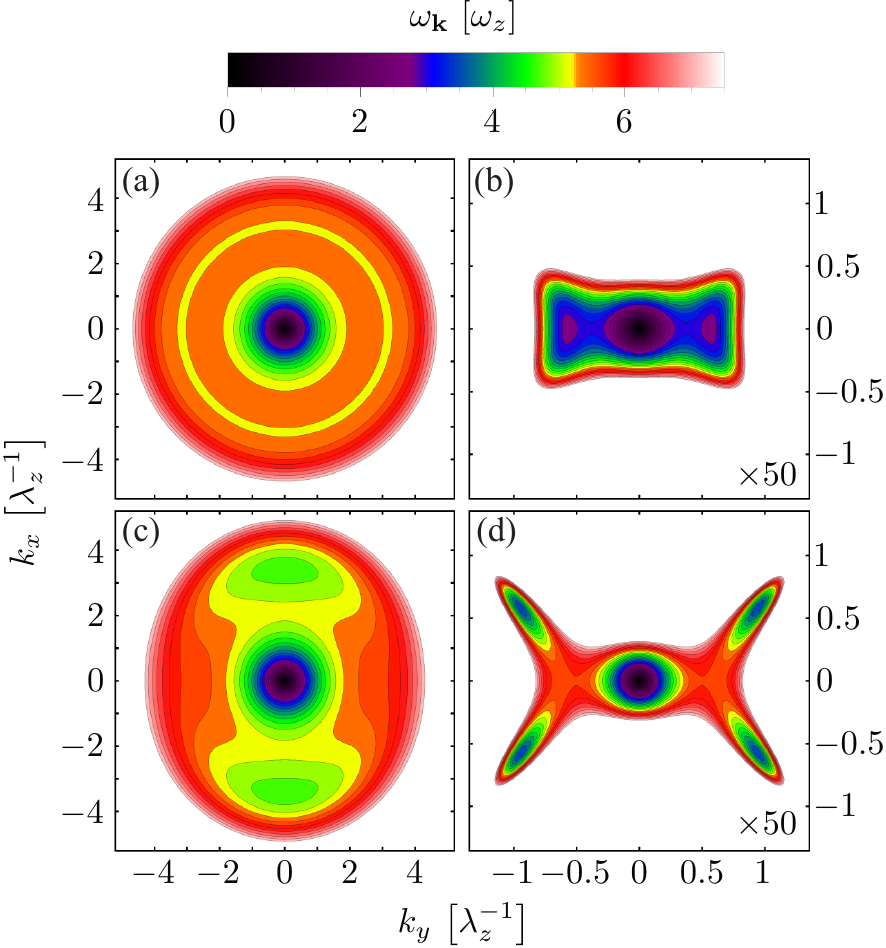}
\caption{(Color online) The dispersion relation for the four cases, in which the rotons are present.
(a) Example of a ring-shaped roton for the special case of $\theta_{F} = 0$, at which the system has rotational symmetry. The density is $n = 17.8\,n_{c}$ and the relative contact interaction is $\eta = 0.2$.
(b) Example of two rotons on $k_{x}$-axis:  $\theta_{F} = 1.55$, $n = 0.57\,n_{c}$ and $\eta = 0.825$.
(c) Expample of two rotons on $k_{y}$-axis: $\theta_{F} = 0.08$, $n = 17.8\,n_{c}$ and $\eta = 0.2$. 
(d) Example of four point-like rotons: $\theta_{F} = 1.55$, $n = 0.8\,n_{c}$ and $\eta = 0.725$.}
\label{fig:rotons}
\end{center}
\end{figure}

Now we identify the rotons that exist in the system and their parameter regime, by determining the number and properties of the dispersion minima. 
We find four different types of stable rotons, cf. \reff{fig:rotons}.

For the special case $\theta_{F} = 0$, for which the system has a rotational symmetry, a ring-shaped roton minimum occurs, as shown in Fig. \ref{fig:rotons} (a). Away from this rotationally symmetric case, the dispersion relation can possess either two or four point-like minima. 
 The two point-like minima can either be on the $k_{x}$-axis, as shown in \reff{fig:rotons} (b), or on the $k_{y}$-axis, \reff{fig:rotons} (c).
 An intriguing case, which is specific to quadrupolar interactions, is the occurrence of four point-like roton minima, as shown in Fig. \ref{fig:rotons} (d).  

\begin{figure}[t!]
\begin{center}
\begin{ruledtabular}
\begin{tabular}{lcc}
Type of rotons	& stable	& unstable\\
none		& \includegraphics[height=.8em]{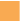}	&	---	\\
circular	& \includegraphics[height=.8em]{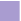}	& \includegraphics[height=.8em]{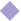}	\\
two		& \includegraphics[height=.8em]{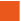}	& \includegraphics[height=.8em]{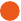}	\\
four		& \includegraphics[height=.8em]{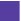}	& \includegraphics[height=.8em]{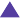}	 \\
\end{tabular}
\end{ruledtabular}
\includegraphics[width=\figurewidth]{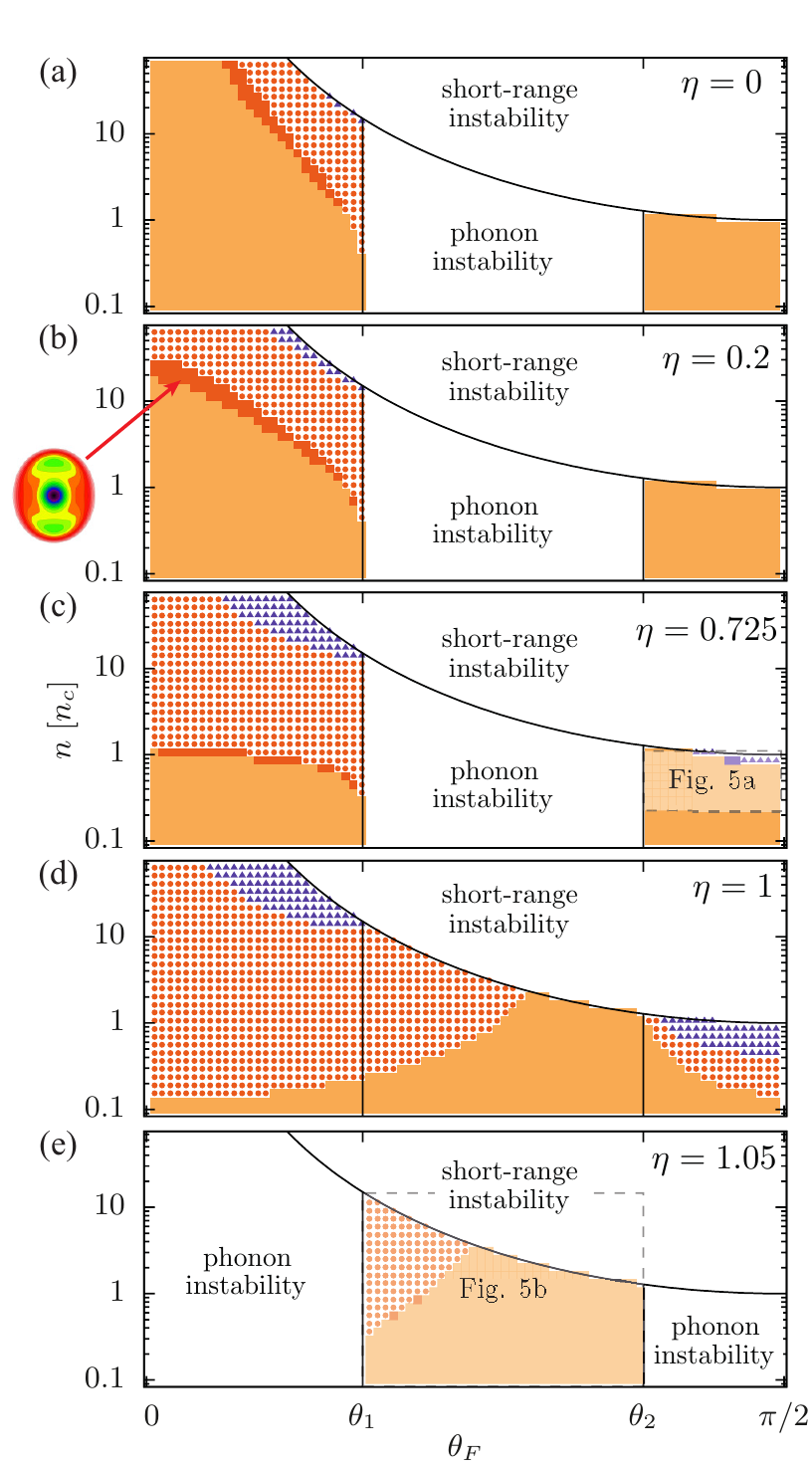}
\caption{(Color online)
The roton properties as a function of the tilting angle $\theta_{F}$ and density $n$. Panels (a) -- (e) correspond to five values of the relative contact interaction $\eta$. 
Note that for fixed $\eta$ the contact interaction $g_{\mathrm{2D}}$ is not a constant, but chosen such that the ratio of \refe{etadef} is kept fixed. For $\theta_{F} < \theta_{1}$ and $\theta_{F} > \theta_{2}$ the contact interaction is attractive, for $\theta_{1} < \theta_{F} < \theta_{2}$ it is repulsive.
A detailed version of the panels (c) and (e) is given in \reff{fig:zoomins}.
}
\label{fig:regime}
\end{center}
\end{figure}

\begin{figure}
\begin{center}
\includegraphics[width=\figurewidth]{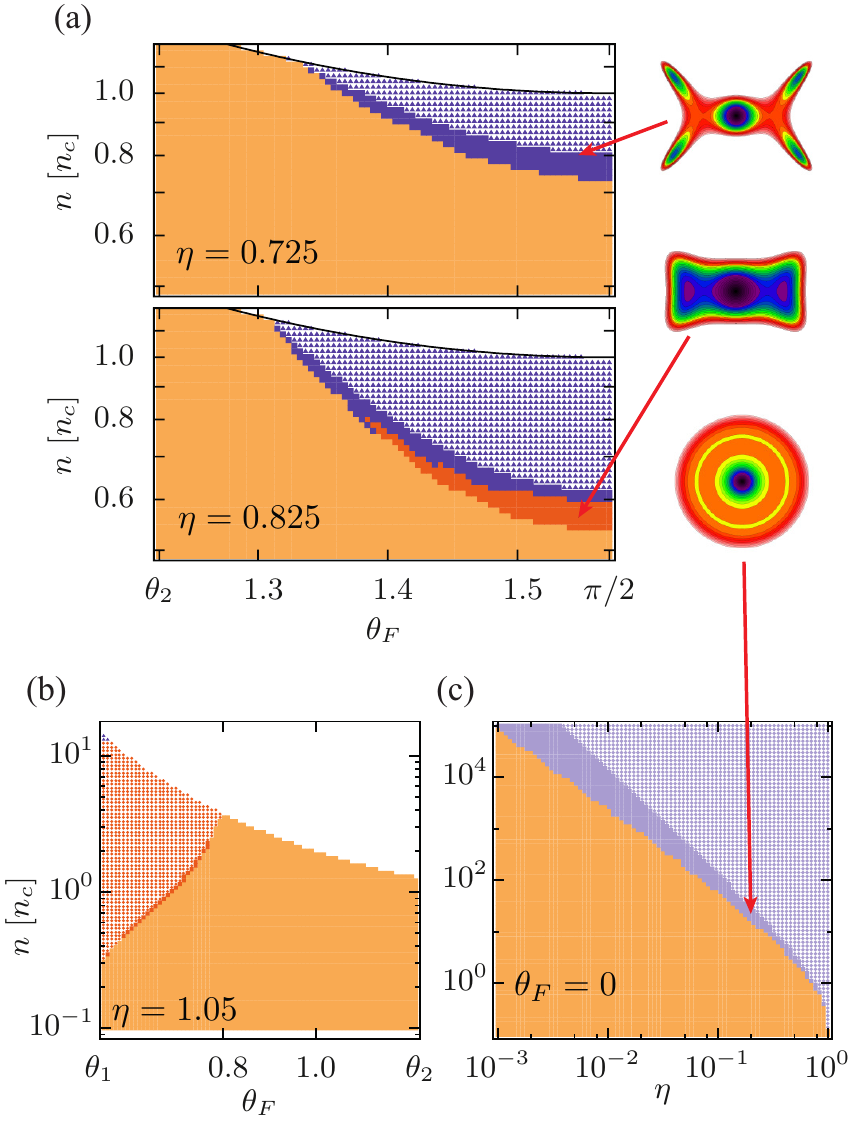}
\caption{(Color online) Detailed view for a few parts of Fig. \ref{fig:regime}.
(a) $\theta_{2} < \theta_{F} < \pi/2$ for $\eta = 0.725$ and $0.825$, which is indicated in \reff{fig:regime} (c). This parameter range contains a regime in which the system has four stable, point-like rotons.
(b) $\theta_{1} < \theta_{F} < \theta_{2}$ for $\eta = 1.05$, which is indicated in Fig. \ref{fig:regime} (e).
(c) $\theta_{F} = 0$, as a function of density $n$ and contact interaction strength $\eta$. Here, the dispersion features a ring-shaped roton minimum. 
}
\label{fig:zoomins}
\end{center}
\end{figure}

In Figs. \ref{fig:regime} and \ref{fig:zoomins} we show the parameter regime in which these types of rotons can occur.

The case of pure QQI and no contact interaction, $\eta = 0$, is shown in Fig. \ref{fig:regime} (a). For large densities, the system displays a short-range instability according to \refe{eq:stabkinf}. For $\theta_{1} < \theta_{F} < \theta_{2}$, the total interaction is attractive for small momenta, leading to the phonon instability in \refe{eq:stabk0}. On the other hand, at small tilting angles, $\theta_{F} < \theta_{1}$, the system shows three regimes: (i) The dispersion is monotonic and has no roton minima for small densities. (ii) Two minima appear on the $k_{y}$-axis as the density is increased. (iii) These rotons become unstable at even larger densities. For large tilting angles, $\theta_{F} > \theta_{2}$, the system is always monotonic, and no rotons are present.
  
We now modify this scenario by turning on a contact interaction. The two cases of $\eta=0.2$ and $\eta = 0.725$ describe a weakly attractive contact interaction for $\theta_{F} < \theta_{1}$ and $\theta_{F} > \theta_{2}$. For $\theta_{F} < \theta_{1}$, the regimes (i)-(ii) move to smaller densities. Furthermore, close to the region of short-range instability, a new regime (iv) of a roton instability in which the dispersion is imaginary for four regions of momentum space occurs.

For sufficiently strong contact interaction, $\eta \geq 0.5$, a new regime occurs for $\theta_{F} > \theta_{2}$, in addition to (i). As the density is increased, the system develops as a new regime, (v) four point-like roton minima for large tilting angles, cf. first panel of \reff{fig:zoomins} (a).  As the density is increased further, these turn into a roton instability (iv).
The axes of the quadrupoles are almost entirely tilted into the plane of the system. While dipolar particles would only be attractive along the dipole axis, quadrupoles have attractive interactions along two directions, both of which are at a non-zero angle to the axis of the quadrupole. If the repulsive parts of the QQI are sufficiently suppressed, this leads to the development of four roton minima, rather than two.
As we show in the second panel of Fig. \ref{fig:zoomins} (a), the regime of stable rotons (v) moves to smaller densities when the contact interaction is increased further. However, as the density is lowered the two pairs of rotons merge into two rotons on the $k_{x}$-axis. Therefore the minimal density to create form stable roton minima in this regime is around $0.6\,n_{c}$.
   
 We now increase $\eta$ further.
 The case of $\eta=1$ is a marginal case, cf. \reff{fig:regime} (d), for which the entire regime of $\theta_{F}$ is stable, and the low-momentum behavior of the dispersion is quadratic instead of linear. In this case, the contact interaction cancels the low-momentum part of the QQI identically.
 For values of $\eta$ larger than $1$, such as $\eta=1.05$ shown in \reff{fig:regime} (e), the regime of the phonon instability is reversed, compared to $\eta<1$. The attractive contact interaction for $\theta_{F} < \theta_{1}$ and $\theta_{F}> \theta_{2}$ is now too large and overcompensates the QQI. However, for $\theta_{1}<\theta_{F}<\theta_{2}$ the contact interaction is now repulsive enough to compensate the attractive QQI and prevent collapse.
 This regime is depicted on a larger scale in \reff{fig:zoomins} (b). We find a large regime with a monotonic dispersion, and a regime with a roton instability of two rotons. Between these two regimes is a small region of stable rotons.
 
 Finally, we show the case of rotational symmetry with $\theta_{F} = 0$ in Fig. \ref{fig:zoomins} (c). As mentioned above, for $\eta >1$ the system is unstable and collapses. For $\eta \leq 1$, three regimes are visible. For smaller densities, the dispersion is monotonic. As the density is increased, the system develops a ring-shaped roton minimum. This minimum becomes unstable, as the density is increased further. The density at the transitions between these regimes depends strongly on the value of $\eta$. Stable rotons for densities near $n_{c}$ are achieved for $\eta$ near $1$. 

\section{Proposed measurement in real space}\label{sec:realspace}
A well-established technique based on two-photon Bragg scattering allows to measure the dynamic structure factor and thereby study the dispersion relation and the roton minima~\cite{Stenger1999,Stamper1999}. In this Section we discuss an alternative scheme, that demonstrates the existence of roton minima in the dispersion and highlights the properties of rotons.
 We consider an experimental setup similar to the one used in Ref.~\cite{Weimer2014} measuring the speed of sound in a stirred BEC.
 During a short time $\Delta{t}$, the system is perturbed with an off-resonant laser beam, which we model as an external potential 
\begin{align}
U_{1}{\left(\vec{r}\right)} &= \frac{V_{0}}{2\pi\sigma^{2}}\exp{\left(-\frac{r^{2}}{2\sigma^{2}}\right)}
\end{align}
with a strength $V_{0}$ and a spatial width $\sigma$. If the system has a linear dispersion at small momenta and is probed with a width, $\sigma$, that is large enough to only probe the low-momentum regime of the dispersion, this perturbation results in an outgoing circular density wave traveling at the speed of sound.
 
 However, for a non-trivial dispersion possessing roton minima, this behavior is modified in a qualitative manner. In particular, the dispersion will necessarily contain regions in which the group velocity is negative. This will result in density waves that propagate towards the location of the perturbation, rather than away from it. Furthermore, the directions of the flow pattern indicate the location and number of roton minima.
 The perturbation term has the form
\begin{align}
\Ham_{1} &= \int{\mathrm{d}{\vec{r}}U_{1}{\left(\vec{r}\right)}\hat{n}{\left(\vec{r}\right)}},\label{Hpert}
\end{align}
where $\hat{n}{\left(\vec{r}\right)}$ is the particle density. We linearize the density $\hat{n}_{\vec{k}}$ in momentum space within the Bogoliubov approximation, which gives $\hat{n}_{\vec{k}} = \sqrt{N_{0}}(u_{\vec{k}}+v_{\vec{k}})(\hat{b}_{-\vec{k}} + \hat{b}_{\vec{k}}^{\dagger})$, where $N_{0}$ is the number of condensed particles. 
 With this expression, \refe{Hpert} is linearized and given by 
\begin{align}\label{eq:Ham1}
\Ham_{1}
&= \sum_{\vec{k}}
S_{\vec{k}}\left(u_{\vec{k}}+v_{\vec{k}}\right)\left(\hat{b}_{-\vec{k}} + \hat{b}_{\vec{k}}^{\dagger}\right).
\end{align} 
Here, $S_{\vec{k}}$ is the Fourier transform of the Gaussian potential, $S_{\vec{k}} = \frac{2\pi V_{0}}{A}\mathrm{e}^{-k^{2}\sigma^{2}/2}$.
  With this term being turned on briefly at time $t=0$, the Bogoliubov operator $\hat{b}_{\vec{k}}{\left(t\right)}$ evolves in time as
\begin{align}\label{eq:ansatz}
\hat{b}_{\vec{k}}{\left(t\right)} &= \hat{b}_{\vec{k}}\mathrm{e}^{-\mathrm{i}\omega_{\vec{k}}t} + A_{\vec{k}}{\left(t\right)}\,,
\end{align}
 where $A_{\vec{k}}{\left(t\right)}$ is zero for $t \leq 0$, and
\begin{align}\label{eq:Akt}
A_{\vec{k}}{\left(t\right)} &= -\frac{\mathrm{i}}{\hbar}S_{\vec{k}}\left(u_{\vec{k}}+v_{\vec{k}}\right)\left(\mathrm{e}^{-\mathrm{i}\omega_{\vec{k}}t}-1\right)\Delta{t}\,
\end{align}
for $t > 0$. We now use this solution for the Bogoliubov operator in the linearized expression for the density, which can be written as $\hat{n}_{\vec{k}} = \hat{n}_{0,\vec{k}}{\left(t\right)} + \hat{n}_{1,\vec{k}}{\left(t\right)}$, where $\hat{n}_{0,\vec{k}}{\left(t\right)}$ is the unperturbed density, and $\hat{n}_{1,\vec{k}}{\left(t\right)}$ is the density perturbation that we are interested in. It is given by
\begin{align}\label{eq:n1k}
\hat{n}_{1,\vec{k}}{\left(t\right)}
&= -\frac{2\pi N_{0}V_{0}\Delta{t}}{m\hbar A}k^{2}\mathrm{e}^{-k^{2}\sigma^{2}/2}\frac{\sin{\left(\omega_{\vec{k}}t\right)}}{\omega_{\vec{k}}}\,.
\end{align}
Using this solution, we construct the density perturbation in real space via $\hat{n}_{1}{\left(\vec{r}\right)} = \sum_{\vec{k}}{\hat{n}_{1,\vec{k}}\mathrm{e}^{-\mathrm{i}\vec{k}\cdot\vec{r}}}$. 
\begin{figure*}
\begin{center}
\includegraphics[width=\textwidth]{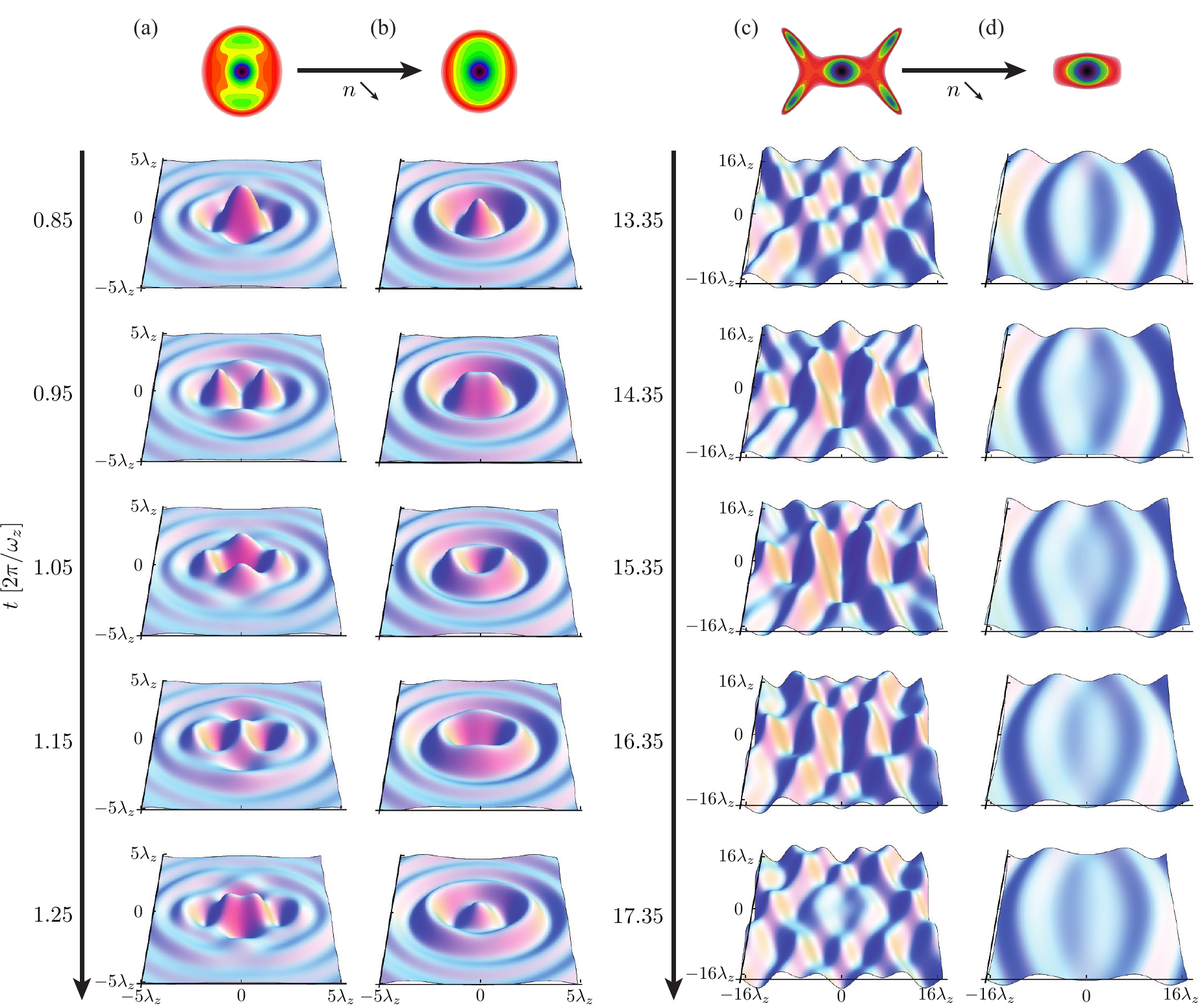}%
\caption{
(Color online) Time evolution of the density in a quadrupolar BEC following a short perturbation at $t=0$ at the origin. 
(a) Example as shown \reff{fig:rotons} (c) with two roton minima on the $k_{y}$-axis: $\theta_{F} = 0.08$, $\eta = 0.2$, $n = 17.8\,n_{c}$. Additional density waves travel along the $y$-axis towards the origin.
(b) Similar configuration as in (a) but with a reduced density, $n = 13.3\,n_{c}$, where no roton minima are present. All density waves are moving outwards from the origin.
(c) Example as shown \reff{fig:rotons} (d) with four roton minima: $\theta_{F} = 1.55$, $\eta = 0.725$, $n = 0.8\,n_{c}$. Density waves form an interference pattern propagating towards the origin.
(d) Similar configuration as in (c) but with a reduced density, $n = 0.6\,n_{c}$, where no roton minima are present. The time evolution of the density consists of outgoing waves only.
}
\label{fig:time}
\end{center}
\end{figure*}

In \reff{fig:time} we show two pairs of examples for this time evolution of the density.
The time sequence in (a) is for the two roton example that was given in \reff{fig:rotons} (c), where $\theta_{F} = 0.08$, $\eta = 0.2$ and $n = 17.8\,n_{c}$.
Panel (b) corresponds to the same values of $\theta_{F}$ and $\eta$, but a reduced density, $n = 13.3\,n_{c}$. We choose the spatial size of the Gaussian perturbation to be $\sigma = 2\,\lambda_{z}$. In the time sequence (a), the density peak at the center initially splits up and moves outwards along the $x$-axis. Later, two peaks appear on the $y$-axis at a similar distance from the origin, however moving inwards. This indicates the occurrence of roton minima on the $k_{y}$-axis for these parameters. For comparison, we show the time sequence (b) where no rotons are present. Here, a density wave propagates outwards in the shape of an elliptic ring, indicating that the dispersion is monotonic.

As the second pair of examples, we show the case of four local rotons, which was given in \reff{fig:rotons} (d), for $\theta_{F} = 1.55$ and $\eta = 0.725$. The density in (c) is $n = 0.8\,n_{c}$ and in (d) it is $n = 0.6\,n_{c}$. The spatial size of the perturbation is $\sigma = 0.6\,\lambda_{z}$.
In the time evolution shown in \reff{fig:time} (c) we now see two incoming density peaks that move toward the $x$-axis, merge and then propagate further towards the origin. The peaks before and after the merging move with different speeds along the axes. This reflects the curvature of the dispersion relation near the roton minima. A large (small) curvature corresponds to a large (small) effective mass which implies that the quasi particles move slower (faster). In other words, the density wave will preferably propagate in the direction of the smallest gradient in the dispersion relation which is not towards the origin but pointing towards the $k_{x}$-axis at an angle. Thus, the density waves created at the roton minima first merge on the $x$-axis, interfere with the outgoing density wave and finally merge at the origin. For the lower density, $n = 0.6\,n_{c}$, no rotons are present and the density waves always propagate outwards, cf. \reff{fig:time}~(d).


\section{Conclusion}\label{sec:conclusion}
We have demonstrated that a quadrupolar two-dimensional condensate can support stable rotonic excitations as well as roton instabilities, which suggest that the system might develop a non-trivial order.  Depending on the alignment angle of the quadrupoles with respect to the system plane, the density, and the magnitude of an additional contact interaction, we identify three types of roton minima. If the quadrupoles are aligned perpendicular to the plane, the roton minimum is ring-shaped, which reflects the rotational symmetry of this state. If the quadrupoles are aligned at a non-perpendicular angle, the dispersion features either two point-like rotons, or, most interestingly, four point-like rotons, which occur for the alignment almost lying within the system plane. Each of these roton types can develop into a roton instability, meaning that the dispersion becomes imaginary at the minimum. 

We study the response of a quadrupolar condensate to a sudden, local perturbation of the density. We demonstrate that there is a qualitative difference in the response of a condensate with a monotonic dispersion and a condensate with a roton minimum. For the monotonic case, the system displays outgoing density waves, whereas the roton minima imply that there are parts of momentum space with negative group velocity. This results in density waves that travel towards the local perturbation rather than away form it. Furthermore, the patterns of these in-flowing density waves indicate which type of roton scenario is present in the system.
These results pave the way to observing exotic roton excitations in the condensates of ultracold homonuclear molecules.


\begin{acknowledgments}
We acknowledge support from the Deutsche Forschungsgemeinschaft through the SFB 925 and the Hamburg Centre for Ultrafast Imaging, and from the Landesexzellenzinitiative Hamburg, which is supported by the Joachim Herz Stiftung.  
\end{acknowledgments}



\onecolumngrid
\appendix


\section{Fourier transform of $U_{\mathrm{2D}}(\vec{r})$}\label{sec:Fouriertrafo}


The quadrupole-quadrupole interaction in a quasi-2D geometry under a tilting $\theta_{F}$ along the $x$-axis is given by
\begin{align}
U_{\mathrm{2D}}{\left(r,\alpha\right)}
&=
-16u_{0}\left(\varrho^{4}+4\varrho^{2}\right)\mathrm{K}_{0}{\left(\frac{\varrho^{2}}{4}\right)}\mathrm{e}^{\frac{\varrho^{2}}{4}}f_{1}{\left(\theta_{F},\alpha\right)}
+8u_{0}\left(\varrho^{4}-2\varrho^{2}+6\right)\mathrm{K}_{0}{\left(\frac{\varrho^{2}}{4}\right)}\mathrm{e}^{\frac{\varrho^{2}}{4}}f_{2}{\left(\theta_{F},\alpha\right)}
\nonumber\\&\qquad
+u_{0}\left(\varrho^{4}+6\varrho^{2}+6\right)\mathrm{K}_{0}{\left(\frac{\varrho^{2}}{4}\right)}\mathrm{e}^{\frac{\varrho^{2}}{4}}f_{3}{\left(\theta_{F},\alpha\right)}
+16u_{0}\left(\varrho^{4}+2\varrho^{2}-2\right)\mathrm{K}_{1}{\left(\frac{\varrho^{2}}{4}\right)}\mathrm{e}^{\frac{\varrho^{2}}{4}}f_{1}{\left(\theta_{F},\alpha\right)}
\nonumber\\&\qquad
-8u_{0}\left(\varrho^{4}-4\varrho^{2}+16-48\varrho^{-2}\right)\mathrm{K}_{1}{\left(\frac{\varrho^{2}}{4}\right)}\mathrm{e}^{\frac{\varrho^{2}}{4}}f_{2}{\left(\theta_{F},\alpha\right)}
-u_{0}\left(\varrho^{4}+4\varrho^{2}\right)\mathrm{K}_{1}{\left(\frac{\varrho^{2}}{4}\right)}\mathrm{e}^{\frac{\varrho^{2}}{4}}f_{1}{\left(\theta_{F},\alpha\right)}
\end{align}
where $r = \left|\vec{r}\right| = \lambda_{z}\varrho$, $\alpha = \arg{\left(\vec{r}\right)}$, $u_{0} = C_{q}/\left(384\sqrt{2\pi}\lambda_{z}^5\right)$ is a constant energy scale, and
$\mathrm{K}_{\nu}{\left(x\right)}$ are the modified Bessel functions of the second kind. Furthermore, we expressed the dependencies on $\theta_{F}$ and $\alpha$ through the following functions:
\begin{subequations}
\begin{align}
f_{1}{\left(\theta_{F},\alpha\right)} &= \sin^{2}{\left(\theta_{F}\right)}\left(7\cos{\left(2\theta_{F}\right)}+5\right)\cos{\left(2\alpha\right)}\\
f_{2}{\left(\theta_{F},\alpha\right)} &= \sin^{4}{\left(\theta_{F}\right)}\cos{\left(4\alpha\right)}\\
f_{3}{\left(\theta_{F},\alpha\right)} &= 20\cos{\left(2\theta_{F}\right)}+35\cos{\left(4\theta_{F}\right)}+9
\end{align}
\end{subequations}
The Fourier-transformed interaction is formally given by
\begin{align}
V_{\mathrm{2D}}{\left(k,\beta\right)}
&=
\int_{0}^{\infty}{r\mathrm{d}{r}\int_{0}^{2\pi}{\mathrm{d}{\alpha}U_{\mathrm{2D}}{\left(\varrho\lambda_{z},\alpha\right)}\mathrm{e}^{-\mathrm{i}p\varrho\cos{\left(\alpha-\beta\right)}}}}\,,
\end{align}
where we introduced the dimensionless quantity $p = \lambda_{z}k$. The angular dependence can be evaluated by integrating the functions $f_{i}{\left(\theta_{F},\alpha\right)}$ over $\alpha$,
\begin{align}
F_{i}{\left(p\varrho,\beta-\varphi_{F}\right)}
&= \int_{0}^{2\pi}{\mathrm{d}{\alpha}f_{i}{\left(\theta_{F},\alpha\right)}\mathrm{e}^{-\mathrm{i}p\varrho\cos{\left(\alpha-\beta\right)}}}\,.
\end{align}
We find
\begin{subequations}
\begin{align}
F_{1}{\left(p\varrho,\beta\right)} &= -2\pi\sin^{2}{\left(\theta_{F}\right)}\left(7\cos{\left(2\theta_{F}\right)}+5\right)\cos{\left(2\beta\right)}\mathrm{J}_{2}{\left(p\varrho\right)}\,\\
F_{2}{\left(p\varrho,\beta\right)} &= 2\pi\sin^{4}{\left(\theta_{F}\right)}\cos{\left(4\left(\beta\right)\right)}\left[\left(1-\frac{24}{\left(p\varrho\right)^{2}}\right)\mathrm{J}_{0}{\left(p\varrho\right)}-\left(\frac{8}{p\varrho}-\frac{48}{\left(p\varrho\right)^{3}}\right)\mathrm{J}_{1}{\left(p\varrho\right)}\right]\,\\
F_{3}{\left(p\varrho,\beta\right)} &= 2\pi\left(20\cos{\left(2\theta_{F}\right)}+35\cos{\left(4\theta_{F}\right)}+9\right)\mathrm{J}_{0}{\left(p\varrho\right)}\,.
\end{align}
\end{subequations}
The modified Bessel functions of the second kind can be defined as $\mathrm{K}_{\nu}{\left(x\right)} = \int_{0}^{\infty}{\mathrm{e}^{-x\cosh{\left(t\right)}}\cosh{\left(\nu t\right)}\mathrm{d}{t}}$. Using a substitution $u^{2} = \cosh{\left(t\right)} - 1$ we find
\begin{subequations}
\begin{align}
\mathrm{e}^{\frac{\varrho^{2}}{4}}\mathrm{K}_{0}{\left(\frac{\varrho^{2}}{4}\right)}
&= \int_{0}^{\infty}{\mathrm{e}^{-\frac{\varrho^{2}}{4}\left(\cosh{\left(t\right)}-1\right)}\mathrm{d}{t}}
= \int_{0}^{\infty}{\frac{2}{\sqrt{u^{2}+2}}\mathrm{e}^{-\frac{\varrho^{2}}{4}u^{2}}\mathrm{d}{u}}\,,\\
\mathrm{e}^{\frac{\varrho^{2}}{4}}\mathrm{K}_{1}{\left(\frac{\varrho^{2}}{4}\right)}
&= \int_{0}^{\infty}{\mathrm{e}^{-\frac{\varrho^{2}}{4}\left(\cosh{\left(t\right)}-1\right)}\cosh{\left(t\right)}\mathrm{d}{t}}
= \int_{0}^{\infty}{\frac{2\left(u^{2}+1\right)}{\sqrt{u^{2}+2}}\mathrm{e}^{-\frac{\varrho^{2}}{4}u^{2}}\mathrm{d}{u}}\,.
\end{align}
\end{subequations}
We introduce an integral of the following form:
\begin{align}
Q_{n,m}{\left(u\right)} &= \int_{0}^{\infty}{\varrho^{n+1}\mathrm{J}_{m}{\left(p\varrho\right)}\mathrm{e}^{-\varrho^{2}u^{2}/4}\mathrm{d}{\varrho}}
= 2^{n+1}p^{m}u^{-\left(m+n+2\right)}\frac{\left(\frac{m+n}{2}\right)!}{m!}{}_{1}\mathrm{F}_{1}{\left(\frac{m+n+2}{2},m+1,-\frac{p^{2}}{u^{2}}\right)}\,,
\end{align}
where the analytic solution is valid for $m\geq0$ and $m+n\geq -1$. Using the recurrence identities of the Bessel functions $\mathrm{J}_{m}$, $\mathrm{J}_{m}{\left(z\right)} = 2\left(m+1\right)\mathrm{J}_{m+1}{\left(z\right)}/z-\mathrm{J}_{m+2}{\left(z\right)}$ and $\mathrm{J}_{m}{\left(z\right)} = 2\left(m-1\right)\mathrm{J}_{m-1}{\left(z\right)}/z-\mathrm{J}_{m-2}{\left(z\right)}$, we find equivalent relations for $Q_{n,m}$,
\begin{subequations}
\begin{align}
Q_{n,m}{\left(u\right)} &= \frac{2\left(m+1\right)}{p}Q_{n-1,m+1}{\left(u\right)}-Q_{n,m+2}{\left(u\right)}\,,\label{eq:ri1}\\
Q_{n,m}{\left(u\right)} &= \frac{2\left(m-1\right)}{p}Q_{n-1,m-1}{\left(u\right)}-Q_{n,m-2}{\left(u\right)}\,.\label{eq:ri2}
\end{align}
\end{subequations}
Furthermore, we define
\begin{align}
P_{n,m} &= \int_{0}^{\infty}{\frac{2}{\sqrt{u^{2}+2}}Q_{n,m}{\left(u\right)}\mathrm{d}{u}}\nonumber\\
&=
\frac{p^{m}}{2^{\mu+m+1}}
\left\{
\frac{\Gamma{\left(\mu +1\right)}\Gamma{\left(2\mu+2\right)}\Gamma{\left(-\frac{\mu+1}{2}\right)}}{\Gamma{\left(m+1\right)}\Gamma{\left(\mu+\frac{3}{2}\right)}} \, {}_2{\mathrm{F}}_2{\left(\mu+1,\mu+1;m+1,\mu+\frac{3}{2};\frac{p^2}{2}\right)}
\right.\nonumber\\
&\qquad\qquad\left.
+\left(-1\right)^{\mu}\frac{\left(2\pi\right)^{3/2} 2^{3\mu}}{p^{2\mu+1}\Gamma{\left(\frac{1}{2}-\mu\right)}\Gamma{\left(m-\mu+\frac{1}{2}\right)}}{}_{2}{\mathrm{F}}_2\left(\frac{1}{2},\frac{1}{2};\frac{1}{2}-\mu ,m-\mu +\frac{1}{2};\frac{p^2}{2}\right)
\right\}\label{eq:Pnm2}
\end{align}
where we set $m+n = 2\mu$ and the solution of the integral is valid for $\mu > -1$. Similar to that, we find
\begin{small}
\begin{align}
R_{n,m} &= \int_{0}^{\infty}{\frac{2\left(u^{2}+1\right)}{\sqrt{u^{2}+2}}Q_{n,m}{\left(u\right)}\mathrm{d}{u}}\nonumber\\
&=
\frac{p^{m}}{2^{\mu+m+1}}
\left\{
\Gamma{\left(2\mu+1\right)}\Gamma{\left(-\frac{\mu-1}{2}\right)}\Gamma{\left(\mu\right)}
\left[
4 \, {}_{2}\tilde{\mathrm{F}}_{2}{\left(\mu ,\mu +1;m+1,\mu +\frac{1}{2};\frac{p^2}{2}\right)}
-2\mu  \,{}_{2}\tilde{\mathrm{F}}_{2}{\left(\mu +1,\mu +1;m+1,\mu +\frac{3}{2};\frac{p^2}{2}\right)}
\right]
\nonumber\right.\\&\left.\qquad
+\left(-1\right)^{\mu}\frac{\left(2\pi\right)^{3/2}2^{3\mu}}{p^{2 \mu +1}}
\left[
{}_{2}\tilde{\mathrm{F}}_{2}{\left(\frac{1}{2},\frac{1}{2};\frac{1}{2}-\mu ,m-\mu +\frac{1}{2};\frac{p^2}{2}\right)}
-\frac{p^2}{2} \,{}_{2}\tilde{\mathrm{F}}_{2}{\left(\frac{1}{2},\frac{3}{2};\frac{3}{2}-\mu ,m-\mu +\frac{3}{2};\frac{p^2}{2}\right)}
\right]
\right\}\,.
\label{eq:Rnm2}
\end{align}
\end{small}%
where we set again $m+n = 2\mu$ and the solution of the integral is valid for $\mu > 0$. Since the integral is linear, the same recurrence identities as for $Q_{n,m}$ apply for $P_{n,m}$ and $R_{n,m}$, respectively. Finally, the Fourier transformed interaction potential can be written as
\begin{align}
V_{\mathrm{2D}}{\left(k,\beta\right)}
&=
32\pi u_{0}f_{1}{\left(\theta_{F},\beta\right)}\left[P_{4,2}+4P_{2,2}-R_{4,2}-2R_{2,2}+2R_{0,2}\right]\nonumber\\
&\quad+
16\pi u_{0}f_{2}{\left(\theta_{F},\beta\right)}\left[P_{4,0}-\left(24 p^{-2}+2\right)P_{2,0}+\left(48p^{-2}+6\right)P_{0,0}+144p^{-2}P_{-2,2}-8p^{-1}P_{3,1}
\right.\nonumber\\
&\left.\qquad
+\left(48p^{-3}+16p^{-1}\right)P_{1,1}-\left(96p^{-3}+48p^{-1}\right)P_{-1,1}
-R_{4,0}+\left(24p^{-2}+4\right)R_{2,0}+8p^{-1}R_{3,1}
\right.\nonumber\\
&\left.\qquad
-\left(48p^{-3}+32p^{-1}\right)R_{1,1}
+\left(96p^{-2}+16\right)R_{0,2}+48R_{-2,4} - 96p^{-1}R_{-1,3}
\right]\nonumber\\
&\quad+
2\pi u_{0}f_{3}{\left(\theta_{F},\beta\right)}\left[P_{4,0}+6P_{2,0}+6P_{0,0}-R_{4,0}-4R_{2,0}\right]\,.
\end{align}
Note, that we made use of the recurrence identities above since not all combinations of $n$ and $m$ fulfill the conditions on the expressions \refe{eq:Pnm2} and \refe{eq:Rnm2} and terms might diverge if considered separately.


\section{Details of the real-space dynamics}\label{sec:Detailed}


In this Section we explain the calculations of Sec.~\ref{sec:realspace} leading to \refe{eq:n1k} in more detail. The annihilation (creation) operator in Fourier space is given by $\hat{a}_{\vec{k}}$ ($\hat{a}_{\vec{k}}^{\dagger}$). Then, the spectral density is given by
\begin{align}
\hat{n}_{\vec{k}}
&= \sum_{\vec{k}^{\prime}}{\hat{a}_{\vec{k}+\vec{k}^{\prime}}^{\dagger}\hat{a}_{\vec{k}^{\prime}}}
=\hat{a}_{\vec{k}}^{\dagger}\hat{a}_{0}+\hat{a}_{0}^{\dagger}\hat{a}_{-\vec{k}}
+ \sum_{\vec{k}^{\prime}\neq0}{\hat{a}_{\vec{k}+\vec{k}^{\prime}}^{\dagger}\hat{a}_{\vec{k}^{\prime}}}\,.
\end{align}
Since we assume a BEC with the occupation number of the condensed mode $\vec{k} = 0$ much larger than the total population of the excited states, $N_{0} \gg \sum_{\vec{k}\neq0}N_{\vec{k}}$, we can (i) replace $\hat{a}_{0}$ and $\hat{a}_{0}^{\dagger}$ by $\sqrt{N_{0}}$ and (ii) neglect the terms which are not at least proportional to $\sqrt{N_{0}}$. Applying the Bogoliubov transformation, following the same arguments as above, we obtain $\hat{n}_{\vec{k}} = \sqrt{N_{0}}\left(u_{\vec{k}}+v_{\vec{k}}\right)\left(\hat{b}_{-\vec{k}} + \hat{b}_{\vec{k}}^{\dagger}\right)$. The perturbation in the Hamiltonian, $\Ham_{1} = \int{\mathrm{d}{\vec{r}}U_{1}{\left(\vec{r}\right)}\hat{n}{\left(\vec{r}\right)}}$, is now expressed in terms of the Fourier representations of density, $\hat{n}{\left(\vec{r}\right)}$, and interaction, $U_{1}{\left(\vec{r}\right)}$, as follows
\begin{align}
\Ham_{1}
&=
\int{\mathrm{d}{\vec{r}}
\sum_{\vec{k}}{V_{1}{\left(\vec{k}\right)}\mathrm{e}^{-\mathrm{i}\vec{k}\cdot\vec{r}}}
\sum_{\vec{k}^{\prime}}{n_{\vec{k}^{\prime}}\mathrm{e}^{-\mathrm{i}\vec{k}^{\prime}\cdot\vec{r}}}
}
\end{align}
Plugging in $V_{1}{\left(\vec{k}\right)} = \int{\mathrm{d}\vec{r}U_{1}{\left(\vec{r}\right)}\mathrm{exp}{\left(\mathrm{i}\vec{r}\cdot\vec{k}\right)}}/A$ and $\hat{n}_{\vec{k}}$, respectively, we find
\begin{align}
\Ham_{1}
&= \sum_{\vec{k},\vec{k}^{\prime}}{
\left(u_{\vec{k}}+v_{\vec{k}}\right)
\left(\hat{b}_{-\vec{k}^{\prime}} + \hat{b}_{\vec{k}^{\prime}}^{\dagger}\right)
}
\frac{V_{0}}{A}\mathrm{e}^{-k^{2}\sigma^{2}/2}
\int{\mathrm{d}{\vec{r}}\mathrm{e}^{-\mathrm{i}\left(\vec{k}+\vec{k}^{\prime}\right)\cdot\vec{r}}}\,.
\end{align}
Making use of the Fourier representation of the $\delta$-distribution, $2\pi\delta{\left(\vec{k}\right)} = \int{\mathrm{d}{\vec{r}}\exp{\left(-\mathrm{i}\vec{k}\cdot\vec{r}\right)}}$, and the fact, that the QQI is mirror-symmetric and thus $\omega_{-\vec{k}} = \omega_{\vec{k}}$, directly leads to \refe{eq:Ham1}. We now solve the equation of motion, 
\begin{align}
\mathrm{i}\hbar\mathrm{d}_{t}{\hat{b}_{\vec{k}}{\left(t\right)}}
&= \left[\hat{b}_{\vec{k}}{\left(t\right)},\Ham_{0}\right] + \left[\hat{b}_{\vec{k}}{\left(t\right)},\Ham_{1}\right]\,,
\end{align}
by inserting the ansatz given in \refe{eq:ansatz}. Using the expressions for the undisturbed Hamiltonian, \refe{eq:Ham0}, the first commutator on the right-hand side becomes
\begin{align}\label{eq:eomH0}
\left[\hat{b}_{\vec{k}}{\left(t\right)},\Ham_{0}\right]
&= \left[\hat{b}_{\vec{k}}\mathrm{e}^{-\mathrm{i}\omega_{\vec{k}}t}+ A_{\vec{k}}{\left(t\right)},
\varepsilon_{0} + \sum_{\vec{k}^{\prime}\neq0}\varepsilon_{k^{\prime}}\hat{b}_{\vec{k}^{\prime}}^{\dagger}\hat{b}_{\vec{k}^{\prime}}\right]
= \mathrm{e}^{-\mathrm{i}\omega_{\vec{k}}t}\sum_{\vec{k}^{\prime}\neq0}\varepsilon_{\vec{k}^{\prime}}\left[\hat{b}_{\vec{k}},\hat{b}_{\vec{k}^{\prime}}^{\dagger}\hat{b}_{\vec{k}^{\prime}}\right]
= \mathrm{e}^{-\mathrm{i}\omega_{\vec{k}}t}\varepsilon_{\vec{k}}\hat{b}_{\vec{k}}\,,
\end{align}
where we applied bosonic commutator relations. Similary, by inserting the perturbation of \refe{eq:Ham1} the second commutator on the right-hand side gives
\begin{align}\label{eq:eomH1}
\left[\hat{b}_{\vec{k}}{\left(t\right)},\Ham_{1}\right]
&= \left[\hat{b}_{\vec{k}}\mathrm{e}^{-\mathrm{i}\omega_{\vec{k}}t} + A_{\vec{k}}{\left(t\right)},
\sum_{\vec{k}^{\prime}}S_{\vec{k}^{\prime}}\left(u_{\vec{k}^{\prime}}+v_{\vec{k}^{\prime}}\right)\left(\hat{b}_{-\vec{k}^{\prime}} + \hat{b}_{\vec{k}^{\prime}}^{\dagger}\right)
\right]
\nonumber\\
&= \mathrm{e}^{-\mathrm{i}\omega_{\vec{k}}t}\sum_{\vec{k}^{\prime}}S_{\vec{k}^{\prime}}\left(u_{\vec{k}^{\prime}}+v_{\vec{k}^{\prime}}\right)\left[\hat{b}_{\vec{k}},
\hat{b}_{-\vec{k}^{\prime}} + \hat{b}_{\vec{k}^{\prime}}^{\dagger}
\right]
= \mathrm{e}^{-\mathrm{i}\omega_{\vec{k}}t}S_{\vec{k}}\left(u_{\vec{k}}+v_{\vec{k}}\right)\,.
\end{align}
However, if we consider the ansatz from \refe{eq:ansatz} directly, we find another expression for the left-hand side of the equation of motion, that is
\begin{align}\label{eq:eom3}
\mathrm{i}\hbar\mathrm{d}_{t}{\hat{b}_{\vec{k}}{\left(t\right)}} &= \varepsilon_{k}\hat{b}_{\vec{k}}\mathrm{e}^{-\mathrm{i}\omega_{\vec{k}}t} + \mathrm{i}\hbar\partial_{t}A_{\vec{k}}{\left(t\right)}\,.
\end{align}
Note, that the first term is equal to the right-hand side of \refe{eq:eomH0}. Thus, the second term must coincide with the right-hand side of \refe{eq:eomH1} resulting in a first order differential equation for $A_{\vec{k}}{\left(t\right)}$, that is
\begin{align}
\partial_{t}A_{\vec{k}}{\left(t\right)} &= -\frac{\mathrm{i}}{\hbar}S_{\vec{k}}\left(u_{\vec{k}}+v_{\vec{k}}\right)\mathrm{e}^{-\mathrm{i}\omega_{\vec{k}}t}\,.
\end{align}
Since we assume only a very short quench within some time interval $\Delta{t}$, we can linearize the integral and find
\begin{align}
A_{\vec{k}}{\left(t\right)} &= -\frac{\mathrm{i}}{\hbar}S_{\vec{k}}\left(u_{\vec{k}}+v_{\vec{k}}\right)\left(\mathrm{e}^{-\mathrm{i}\omega_{\vec{k}}t}-1\right)\Delta{t}
\end{align}
We chose the integration constant in such a way, that the boundary condition $A{\left(t=0\right)} = 0$ and thus $\hat{b}_{\vec{k}}{\left(t=0\right)} = \hat{b}_{\vec{k}}$ is fulfilled. This is the solution given by in \refe{eq:Akt}.


\end{document}